\documentclass{PoS}
\pdfoutput=1

\usepackage{wrapfig}
\usepackage{booktabs}
\usepackage{amsmath}
\usepackage{amsbsy}
\usepackage{amssymb}

\let\OLDthebibliography\thebibliography
\renewcommand\thebibliography[1]{
  \OLDthebibliography{#1}
  \setlength{\parskip}{0pt}
  \setlength{\itemsep}{0pt plus 0.3ex}
}

\def\aap{{Astronomy and Astrophys.}}

\def\grl{{\it Geophys.~Res.~Lett.\ }}
\def\apj{{\it Astrophys. J.\ }}
		
\def\apjl{{\apj\ \it Lett.\ }}

\def\pre{{\it Phys.~Rev.~E\ }}

\def\prd {{\it Phys. Rev. D}}

\DeclareMathAlphabet\mathbfcal{OMS}{cmsy}{b}{n}
\DeclareMathAlphabet\mathcal{OMS}{cmsy}{n}{n}

\newcommand{\mb}[1]{\mathbf{#1}}

\title{Anomalies in Cosmic Ray Composition: Explanation Based on Mass to Charge Ratio}

\ShortTitle{Anomalies in Cosmic Ray Composition: Explanation Based on Mass to Charge Ratio}

\author{\speaker{Adrian Hanusch}\\%
        Rostock University, Germany\\
        E-mail: \email{adrian.hanusch@uni-rostock.de}}
        
\author{Tatyana Liseykina\\
        Rostock University, Germany\\
        E-mail: \email{tatyana.liseykina@uni-rostock.de}}
        
\author{Mikhail Malkov\\
        University of California, San Diego, USA\\
        E-mail: \email{mmalkov@ucsd.edu}}

\abstract{High precision spectrometry of galactic cosmic rays (CR) has revealed 
the lack of our understanding of how different CR elements are extracted from 
the supernova environments to be further accelerated in their shocks. Comparing 
the spectra of accelerated particles with different mass to charge ratios is a 
powerful tool for studying the physics of particle injection into the diffusive 
shock acceleration (DSA). Recent AMS-02 demonstration of the similarity of He/$p$, 
C/$p$, and O/$p$ rigidity spectra has provided new evidence that injection is a 
mass-to-charge dependent process. We performed hybrid simulations of collisionless 
shocks and analyzed a joint injection of $p$ and He$^{2+}$ in conjunction with 
upstream waves they generate. By implication, our results equally apply to C and O 
fully ionized ions, since they have similar mass to charge ratios. By convolving 
the time-dependent injection rates of $p$ and He, obtained from the simulations, 
with a decreasing shock strength over the active life of SNRs, we generate the 
integrated SNR spectra for $p$ and He. These spectra are consistent with the AMS-02 
and Pamela data and earlier theoretical predictions. Our interpretation of the 
elemental anomaly is therefore intrinsic to collisionless shock mechanisms and does 
not require additional assumptions, such as the contributions from several different 
SNRs, their inhomogeneous environments or acceleration from grains.}

\FullConference{35th International Cosmic Ray Conference\\
                 10-20 July, 2017\\
                 Bexco, Busan, Korea}

\begin{document}

\section{Introduction\label{sec:Introduction}}

\begin{wrapfigure}{O}{0.42\textwidth}%
\vspace{-3pt}
\includegraphics[scale=0.55]{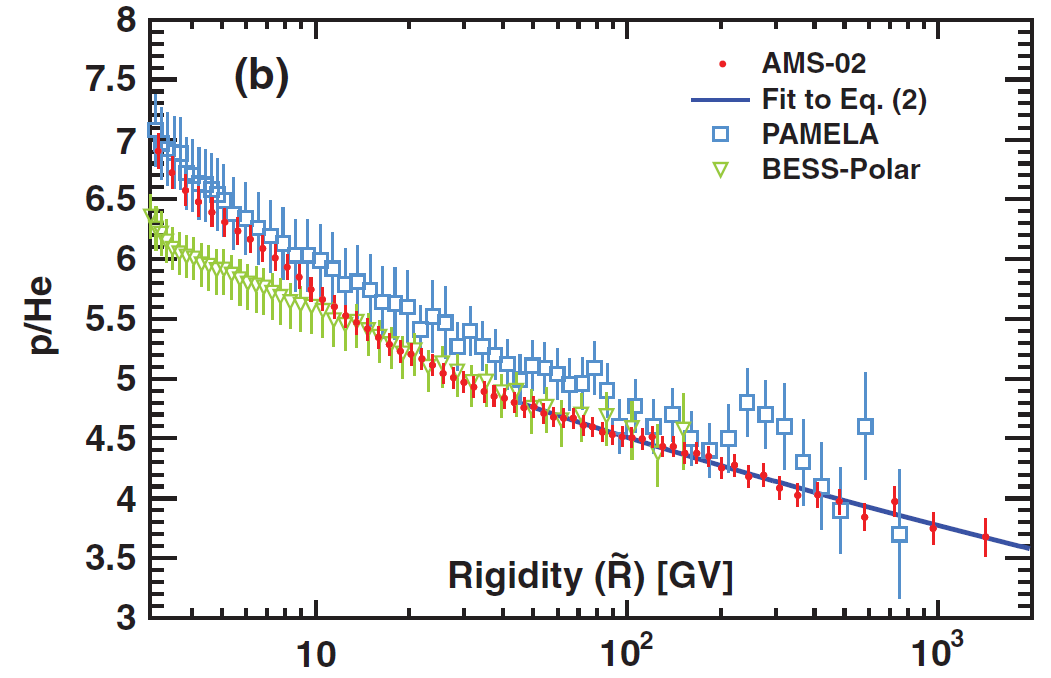}
\vspace{-7pt}
\caption{The $p$/He ratio as a function of particle rigidity. The plot is adopted
from \cite{AMS02He2015PhRvL}.\label{fig:The-p/He-ratio}}
\vspace{-5pt}
\end{wrapfigure}

It is now well documented that the rigidity spectral indices of protons and Helium are different by $\approx0.1$ \cite{Adriani11,AMS02He2015PhRvL}, Fig.~\ref{fig:The-p/He-ratio}. The scaling shown in this plot is likely
to continue to higher rigidities, according to other observations
\cite{CREAM11}. These findings call into question the leading hypothesis
of cosmic ray (CR) origin. According to this hypothesis, they are
accelerated out of an interstellar plasma of the Milky Way when it
is swept by blast waves of supernovae. 

The CR acceleration mechanism is believed to be electromagnetic
in nature. Particles gain energy while being scattered by converging
plasma flows upstream and downstream of a supernova remnant (SNR)
shock. The mechanism was originally proposed in 1949 by Fermi \cite{Fermi49}
and actively researched under the name diffusive shock acceleration (DSA).
It is not difficult to see that even a small, e.g., a $\approx0.1$,
difference in the rigidity spectral indices of different elements
may seriously undermine any (not only the DSA!) electromagnetic acceleration
mechanism. It is sufficient to write the particle equations of motion
in terms of their rigidity, $\mathbfcal R=\mathbf{p}c/eZ$, instead
of momentum $\mathbf{p}$ ($Z$ is the charge number):

\begin{minipage}[c][60pt]{0.5\textwidth}
  \vspace{2pt}
  \begin{equation}
    \frac{1}{c}\frac{d\mathbfcal R}{dt} = \mathbf{E}\left(\mathbf{r},t\right)+\frac{\mathbfcal R\times\mathbf{B}\left(\mathbf{r},t\right)}{\sqrt{\mathcal{R}_{0}^{2}+\mathcal{R}^{2}}},
    \label{eq:RigMotion}
  \end{equation}
\end{minipage}
\begin{minipage}[c][60pt]{0.5\textwidth}
  \vspace{2pt}
  \begin{equation}
    \frac{1}{c}\frac{d\mathbf{r}}{dt} = \frac{\mathbfcal R}{\sqrt{\mathcal{R}_{0}^{2}+\mathcal{R}^{2}}}.
    \label{eq:CoordMotion}
  \end{equation}
\end{minipage}

\noindent
The electric, $\mathbf{E}\left(\mathbf{r},t\right)$, and magnetic,
$\mathbf{B}\left(\mathbf{r},t\right)$, fields here are 
arbitrary. So, the equations apply to the CR acceleration in an SNR shock,  propagation through the turbulent interstellar medium (ISM), and eventual escape from the Milky Way. It follows
that all species with $\mathcal{R}\gg\mathcal{R}_{0}=Am_{p}c^{2}/Ze$
($A$ is the atomic number and $m_{p}$- proton mass, so $\mathcal{R}_{0}\simeq A/Z$
GV) have nearly identical orbits in the phase space $\left({\bf r},\mathbfcal R\right)$.
If, say, protons and helium atoms are already accelerated to $\mathcal{R=}\mathcal{R}_{1}\gg\mathcal{R}_{0}$
with a stationary $p$/He ratio, this same ratio will persist for all
$\mathcal{R}\gg\mathcal{R}_{0}$, in apparent contradiction with Fig.~\ref{fig:The-p/He-ratio}. 

Three different ideas have been entertained to explain the above paradox:
(1) contributions from several SNRs with different $p$-He mixes and
spectral slopes; (2) CR spallation in the ISM that introduces particle
sources and sinks in Eq.~(\ref{eq:RigMotion}); (3) shock evolution
in time. The latter is the case, but this scenario is difficult to
defend. Assuming the $p$/He ratio fixed at some fiducial rigidity $\mathcal{R}=\mathcal{R}_{1}\gg\mathcal{R}_{0}$,
it must increase while the shock strength naturally decreases. While
not impossible, the $p$/He increase must occur at a rate consistent
with the observed $p$/He slope in rigidity. The crucial point here is
that the power-law index of shock-accelerated particles, $q,$ decreases
with the Mach number $M(t)$ in a fully determined way: 
\begin{equation}
q=-d\ln f/d\ln\mathcal{R}=4/\left(1-M^{-2}\right),\label{eq:qOfM}
\end{equation}
where $f$ is the CR distribution function. Therefore, to reproduce
the $0.1$ index difference between $p$ and He correctly, the $p$/He
ratio at $\mathcal{R}=\mathcal{R}_{1}\gg\mathcal{R}_{0}$
must depend on the shock Mach number in a specific way. This dependence,
in turn, is an intrinsic property of collisionless shock and cannot
be adjusted to fit the data, thus making the scenario (3) fully testable.

Unlike the scenario (3) above, (1) is not testable because the individual
properties of contributing sources are unknown. Besides, it will likely
fail the Occam's razor test, especially after the AMS-02 has measured
$p$/C and $p$/O ratios to be identical to those of $p$/He \cite{AMS-02PrRel2016}.
It was also pointed out in \cite{2017arXiv170305772M} that it would
be impossible to maintain the spectral slopes in these ratios nearly
constant over an extended rigidity range. Meanwhile, according to
\cite{VladimirMoskPamela11}, the spallation effects (2) cannot fully
explain the $p$/He rigidity dependence, either. It follows that the
time dependence of the subrelativistic acceleration phase (injection
into DSA, option 3 above) is the most realistic scenario to consider. 

The time dependence of particle acceleration at an SNR shock comes
in two flavors. Firstly, the medium into which the shock propagates
may be inhomogeneous \cite{Ohira11} (effect of SNR environment).
If also the background $p$/He ratio increases outward, this ratio will
decrease with rigidity after the acceleration, since higher rigidities
are dominated by earlier times of acceleration history when the He
contribution was higher. Secondly, the shock weakening (decrease of
the shock Mach number) also makes the acceleration time dependent.
The primary problem with the environmental explanation is that not
only the He concentration must be \emph{assumed} to decrease with
growing shock radius at a specific rate (one free parameter), but
also the carbon and oxygen concentrations must decrease at the same
rate (two more free parameters). This conclusion follows from the
C/He and O/He flux ratios being independent of rigidity \cite{AMS-02PrRel2016}.
So, He, C, and O are likely to share their
acceleration and propagation history. One corollary of this is that
C and O are unlikely to be pre-accelerated from grains, contrary to
some earlier suggestions \cite{Ohira2016PhRvD}. Moreover, the equivalence
between the He, C and O spectra corroborates the conclusion \cite{VladimirMoskPamela11}
that spallation effects are insufficient to account for the observed
differences in rigidity spectra between $p$ and elements whose $A/Z$
values are similar but higher than that of the protons. Note that
it is crucial to use the rigidity dependence of the \emph{fractions} 
of different species as a primary probe into the intrinsic properties
of CR accelerators. Unlike the individual spectra, the fractions are
unaffected by the CR propagation, reacceleration, and losses from
the galaxy, as long as the spallation is negligible. 

Besides being in tension with the AMS-02 recent results, the mechanisms
considered so far require special conditions, such as how the CR sources
are distributed in the ISM, types of supernova progenitors, or their
environments. The question is then if a shock can modify an element
abundance at higher energies by selectively accelerating it out of
a homogeneous background plasma with no additional assumptions? It
has been argued in \cite{m98} that such elemental selectivity of
the initial phase of the DSA (injection) must, indeed, occur in quasi-parallel
shocks. The efficiency of injection mechanism was shown to have the
following properties: it depends on the shock Mach number;
its efficiency increases with $A/Z$, then saturates at a level that
grows with $M$. These properties have been established by analytic
calculations of ion injection into the DSA \cite{mv95,m98,MV98AdSpR..21..551M}. 

The publication of high-precision measurements of $p$/He ratio by the
Pamela collaboration \cite{Adriani11}, prompted the authors of \cite{MDSPamela12}
to apply the analytic injection theory to the case $A/Z=2$ (specifically
to He$^{2+}$, also valid for fully stripped C and O, accurately measured
later by AMS-02). These calculations produced an excellent fit to
the Pamela data in the relevant rigidity range $2<\mathcal{R}<200$GV.
Note that lower rigidities are strongly affected by solar modulation,
while at higher rigidities the Pamela statistics was insufficient
to make a meaningful comparison. 

\section{Analysis\label{sec:Analysis}}

The purpose of this paper is to demonstrate that the recent high-precision
measurements of elemental spectra with different $A/Z$ are not only
consistent with the hypothesis of CR origin in the SNR, but also strongly
support it. Although a similar stand has been taken in \cite{MDSPamela12}
about the Pamela findings \cite{Adriani11}, the new AMS-02 data and
recent progress in shock simulations allow us to establish crucial
missing links in the CR-SNR relation. In particular, the coincidence
in accelerated particle spectral slopes of now three different elements
with $A/Z\simeq2$ (He, C, and O) 
points to an intrinsic, $A/Z$-based element selection mechanism
and rules out incidental ones, such as those based on inhomogeneous, incompletely ionized, 
or dusty shock environments. Note
that the latter mechanism was justified by integrated element
abundance, whereas the individual rigidity spectra
have become known only now. 

On the theoretical side, the $p$/He calculations in \cite{MDSPamela12}
have been based on an analytic theory that allows some freedom in
choosing seed particles for injection \cite{mv95}. These aspects
have been recently discussed, along with the limitations of hybrid
simulations in \cite{2017arXiv170305772M}. Nevertheless, simulations
can remove many uncertainties in the analysis, thus greatly improving
our understanding of the $A/Z$ selectivity. This will be clear after
the following short discussion of self-regulation and elemental selectivity
of injection. 

Injected protons drive unstable Alfv\'en waves in front of the shock.
These waves control the injection of all particles by regulating their
shock crossing needed to gain energy. Proton-driven waves trap them most efficiently. 
Furthermore, the waves are almost frozen into the local fluid so, when crossing
the shock interface, they trap most particles and prevent them from
escaping upstream, thus significantly reducing their odds for injection.
Again, most efficient is namely the proton trapping, while, e.g.,
He$^{2+}$ ions have better chances to escape upstream and be injected. 
The trapping becomes naturally stronger with growing wave amplitude that also grows with
the Mach number. This trend is more pronounced for the protons than
He, which is crucial for the injection selectivity. 

Another limitation that we remove using hybrid simulations concerns treating He ions
as test particles. Several hybrid simulations, addressing the
acceleration efficiency of alpha particles, did include them self-consistently
\cite{BurgessHe1989GeoRL..16..163B,TrattnerScholerHep91inj} while
other considered them as test particles, e.g. \cite{Caprioli2017pHe}.
The fully self-consistent simulations, however, do not provide sufficiently
detaled Mach number scans of the $p$/He injection ratio 
to test the injection bias mechanism discussed earlier. Although
the test particle approximation is often considered to be sufficient
because of the large ($\simeq10$) $p$/He density ratio, the He ions
drive resonant waves that are typically two times longer than the
waves driven by the protons. In the wave-paticle interaction, the
resonance condition is often more important than the wave amplitude.
In adition, the rational relation between the respective wave lengths
(2:1) is suggestive of parametric interactions between them. Such
interaction should facilitate a cascade to longer waves which are
vital for the DSA, not just for particle injection. To conclude this
secion, a state-of-the-art self consistent simulation is required
to test the theoretically predicted $p$/He injection bias.

\section{Simulations\label{sec:Simulation}}

We investigate particle injection into the DSA using hybrid simulations,
whereby electrons are treated as a massless fluid. This technique
is justified because the relevant scales are determined by the ions.
The electron fluid is described by the following equation

\vspace{-7pt}
\begin{equation}
0=-e\,n_{e}\left(\mb{E}+\frac{1}{c}\,\mb{v}_{e}\times\mb{B}\right)-\nabla p_{e}+e\,n_{e}\,\eta\,\mb{J},\label{eq:massless-fluid}
\end{equation}
where $-e$, $n_{e}$, and $\mb{v_{e}}$, are the electron charge,
density and bulk velocity. $\mb{E}$, $\mb{B}$, and $\mb{J}$ are
the electric field, magnetic field, and total current. The electron
pressure $p_{e}$ and the resistivity $\eta$ are both assumed to
be isotropic, i.e., scalar quantities. The electron pressure is related to their density, $n$, 
by an adiabatic equation of state, $p_{e}\sim n^{\gamma_{e}}$, with the index
$\gamma_{e}=5/3$. The ions are treated kinetically as macro-particles,
whose motion is governed by Newton's equations

\vspace{-7pt}
\begin{equation}
  m_{i}\,\frac{d\mb{v}}{dt} = q_{i}\left(\mb{E}+\frac{1}{c}\,\mb{v}\times\mb{B}-\eta\,\mb{J}\right)
  \qquad \text{and} \qquad 
  \frac{d\mb{x}}{dt} = \mb{v},
  \label{eq:ions-motion}
\end{equation}
where $m_{i}$ and $q_{i}$ are the ion mass and charge, respectively.
Eqs.~(\ref{eq:massless-fluid}) and (\ref{eq:ions-motion}) 
are non-relativistic, as $|\mathbf{v}| \ll c$ holds during the injection phase. The
electric field is calculated from Eq.~(\ref{eq:massless-fluid})
using the ion density $n_{i}$ and current $\mb{J}_{i}$ obtained
from particle orbits. The total current is, in turn, obtained from
Ampere's law in magnetostatic approximation, while the magnetic field 
evolves according to Faraday's law

\vspace{-7pt}
\begin{equation}
  \nabla\times\mb{B}=\frac{4\pi}{c}\,\mb{J}
  \qquad \text{and} \qquad
  \frac{1}{c}\,\partial_{t} \, \mb{B}=-\nabla\times\mb{E}.
  \label{eq:faraday}
  \vspace{-3pt}
\end{equation}
In the simulations, lengths are given in units of $c/\omega_{p}$,
with $\omega_{p}=\sqrt{4\pi\,n_{0}\,e^{2}/m_{p}}$ being the proton
plasma frequency, where $n_{0}$ is the upstream density and $e$
and $m_{p}$ are the charge and the mass of the proton, respectively.
Time is measured in the units of inverse proton gyrofrequency, $\omega_{c}^{-1}$,
with $\omega_{c}=e\,B_{0}/m_{p}\,c$. Here $B_{0}$ is the magnitude
of the background magnetic field, and the velocity is normalized to
the Alfv\'en velocity $v_{A}=B_{0}/\sqrt{4\pi\,n_{0}\,m_{p}}$.

\begin{wraptable}{O}{0.4\textwidth}%
\vspace{-20pt}
 \centering 
 \def\arraystretch{0.85}
\begin{tabular}{ccc}
\toprule 
$v_{0}/v_{A}$  & $L_{x}/(c/\omega_{p})$  & $t_{\mathrm{max}}/\omega_{c}^{-1}$ \tabularnewline
\midrule 
3  & 7200  & 3500 \tabularnewline
4  & 7200  & 2450 \tabularnewline
5  & 7200  & 1950 \tabularnewline
7  & 9600  & 1750 \tabularnewline
10  & 9600  & 1700 \tabularnewline
15  & 9600  & 1500 \tabularnewline
20  & 12000  & 1000 \tabularnewline
25  & 14400  & 1000 \tabularnewline
30  & 17280  & 750 \tabularnewline
\bottomrule
\end{tabular}
\vspace{-5pt}
\caption{Size of the simulation box and duration of the simulation for different
initial velocities $v_{0}$. For the low-$M$ shocks it takes longer
until the energy spectra are converged, hence longer runs are necessary.}
\label{tab:sim} %
\vspace{-5pt}
\end{wraptable}%

In our simulations we use a realistic composition of the plasma consisting
of 90\% $p$ and 10\% He$^{2+}$ in number density. Hence, the
helium ions cannot be regarded as test particles. Furthermore, 
the simulations are one-dimensional
spatially, but all three components of the fields and velocity are
included. The reduced number of dimensions allows us to increase the
number of waves and particle statistics, both being crucial for an
adequate description of the downstream thermalization process. It
requires a well-developed wave turbulence to ensure strong wave-particle
interactions with extended resonance overlapping. These conditions
are key to the entropy production, e.g. \cite{ZaslavHamChaos07},
not easy to meet in simulations. They are arguably more important
than, e.g., possible shock rippling effects, not captured by 1D simulations.
At the same time these, and other phenomena occurring at the shock
ramp cannot be accurately characterized within hybrid simulations
anyway as they require a full kinetic treatment \cite{Liseykina15,MSetal_IAshocks2016}
(for further recent discussion of the relevant simulation aspects
see \cite{2017arXiv170305772M}).

The shock simulation is initiated by sending a supersonic and superalfv\'enic
plasma flow with velocity $v_{0}>c_{s}\approx v_{A}$ against a reflecting
wall. The shock forms due to the interaction of the emerging counter-propagating
streams. The background magnetic field is set parallel to the shock
normal $\mathbf{B}_{0}=B_{0}\mathbf{x}$. The upstream plasma betas
are set to $\beta_{e}=\beta_{i}=1$. The simulation box has a length
of $7200-17280\,c/\omega_{p}$, depending on the initial velocity $v_{0}$
(see Tab.~\ref{tab:sim}). The grid spacing is $\Delta x=0.2\;c/\omega_{p}$
with 100 particles per cell for each species and the time step is
chosen to be $\Delta t=0.01/(v_{0}/v_{A})\;\omega_{c}^{-1}$. All numerical
parameters have been checked for convergence.

\section{Results}\label{sec:Results}

\subsection{injection efficiency}

\begin{wrapfigure}{r}{0.5\textwidth}%
\vspace{-25pt}
 \includegraphics[width=0.5\textwidth]{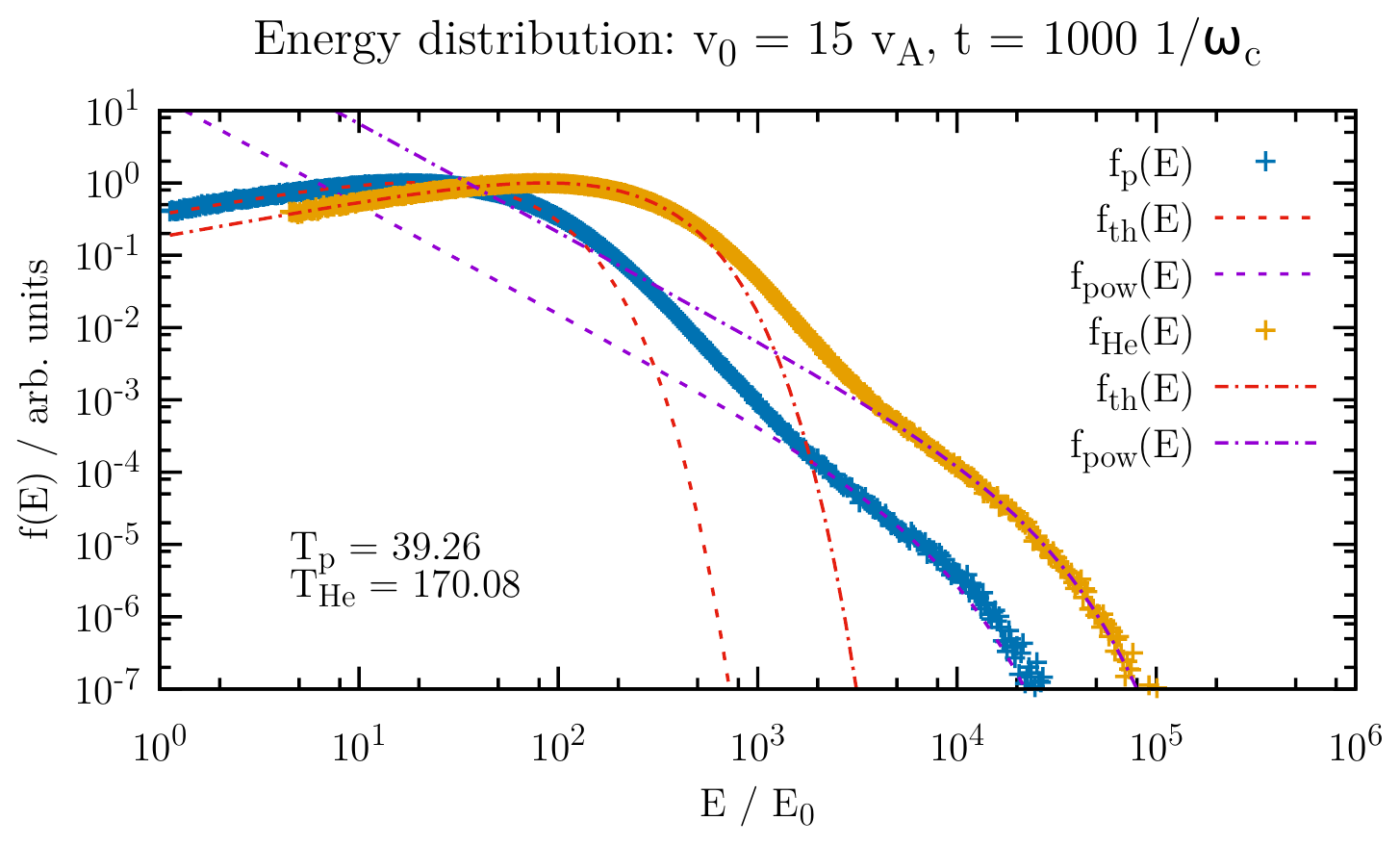}
\vspace{-25pt}
\caption{Downstream energy spectrum of protons (blue) and He$^{2+}$ ions (orange)
for a simulation with $v_{0}=15v_{A}$. At low $E$ the Maxwellian
distribution is visible. For energies $E\gg T$ the distribution follows
a power-law. \label{fig:energy_dist}}
\vspace{-15pt}
\end{wrapfigure}

To investigate the elemental selectivity of the injection mechanism,
we determine the injection efficiency of each particle species ($p$
and He$^{2+}$) included in the simulation.
We obtain the energy spectra of all particles downstream of the shock
transition from the simulation using a logarithmic binning procedure.
Energy is measured in units of $E_{0}=\frac{1}{2}m_{p}\,v_{A}^{2}$.
In general, the spectra of both particle species exhibit the same
main features\textemdash{}a Maxwellian distribution and a power-law
tail (see Fig.~\ref{fig:energy_dist})\textemdash{}where the He$^{2+}$
energy spectrum is shifted towards higher energies. This shift is
a result of the higher He mass under velocity (not energy) randomization
occurring upon crossing the shock. Additionally, a contribution of
supra-thermal particles obscures the transition from the Maxwellian
to the power-law making it difficult to determine the injection efficiency
directly. Hence, we fit a thermal distribution $f_{\mathrm{th}}=a\,E^{1/2}\,\exp(-E/T)$,
where $T$ is the downstream temperature of the plasma as well as
a power-law with a cut-off, $f_{\mathrm{pow}}=b\,E^{-q}\,\exp(-E/E_{\mathrm{cut}})$,
where $E_{\mathrm{cut}}$ is the cut-off energy. 

As expected, the
temperature of the He$^{2+}$ ions is approximately four times the
proton temperature and the time needed for thermalization is longer
for the heavier ions. The injection efficiency is then calculated
using the fits mentioned earlier, as 
\begin{equation}
\eta_{\mathrm{inj}}=\frac{f_{\mathrm{th}}(E_{\mathrm{inj}})}{\int_{0}^{\infty}f_{\mathrm{th}}(E)\;\mathrm{d}E}\label{eq:inj-efficiency}
\end{equation}
with $E_{\mathrm{inj}}$ defined from $f_{\mathrm{th}}(E_{\mathrm{inj}})=f_{\mathrm{pow}}(E_{\mathrm{inj}})$.

\begin{wrapfigure}{O}{0.5\textwidth}%
\vspace{-5pt}
\includegraphics[width=0.5\textwidth]{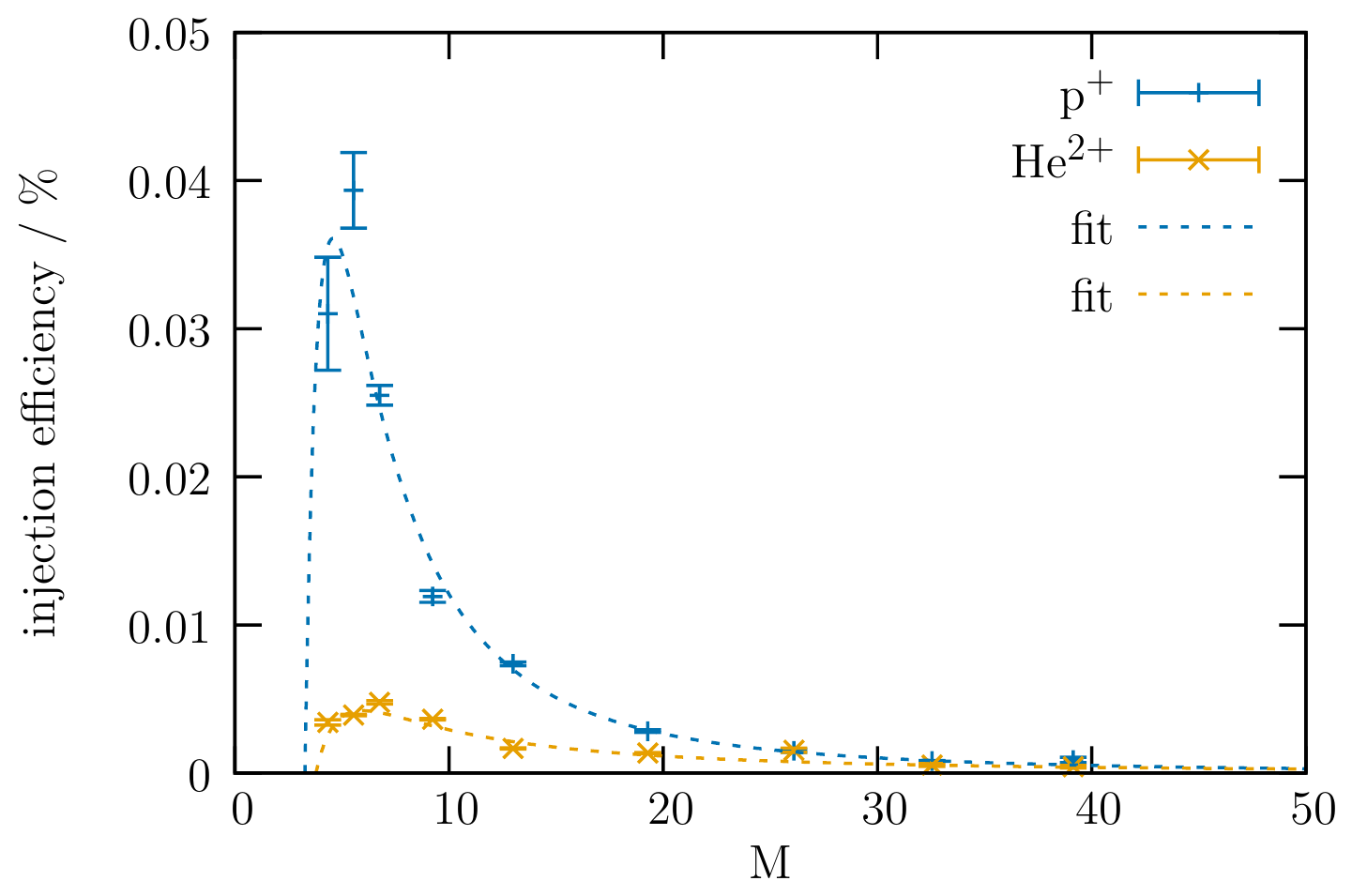}
\vspace{-20pt}
\caption{ Values for the injection efficiency of protons (blue) and He$^{2+}$
ions (orange) obtained from the simulation according to Eq.~(\ref{eq:inj-efficiency})
for different shock velocities. \label{fig:inj-efficiency}}
\vspace{-5pt}
\end{wrapfigure}

We have performed a number of simulations for a range of different
initial upstream flow velocities $v_{0}$ (see Tab.~\ref{tab:sim}),
corresponding to different shock Mach numbers, $M=(v_{0}+v_{sh})/v_{A}$,
where $v_{sh}$ is the shock velocity in the downstream (simulation)
rest frame. After the energy spectra are converged, we apply our method
of calculating the injection efficiency at multiple times in a time
interval of $250\,\omega_{c}^{-1}$ and compute the average. The resulting
values of $\eta_{\mathrm{inj}}^{\alpha}(M)$ are depicted in Fig.~\ref{fig:inj-efficiency}.

In general, the injection efficiency as a function of the shock Mach
number exhibits a similar shape for $p$ and He$^{2+}$. It increases
for $M\lesssim5$ for $p$, while for He$^{2+}$ an increase up
to $M\lesssim7$ is visible. For higher $M$ the injection efficiency
decreases for both species. However, the injection of protons dominates
for low Mach number shocks with $\eta_{\mathrm{inj}}^{p}$ exceeding
the value of $\eta_{\mathrm{inj}}^{\mathrm{He}}$ by an order of magnitude.
Furthermore, the maximum of the injection efficiency of $p$ shifts 
towards smaller $M$ compared to He$^{2+}$. At larger $M$, the
injection efficiency shows the predicted behavior $\eta_{\mathrm{inj}}(M)\sim\ln(M/M^{*})/M$
(Eq.~(76) in \cite{m98}). The prevalence of proton injection at
slow shocks is also noticeable in the downstream temperature ratio
$T_{\mathrm{He}}/T_{p}$, which exceeds the expected ratio of four
for $M<15$. This can be explained by the larger fraction of energy
which is converted into the energy of the accelerated particles for
the protons \cite{Gieseler00}.

\subsection{Proton-to-helium ratio}

Our aim is now to combine the Mach number dependent injection efficiency
with the time dependence of the evolution of a SNR in order to model
the time-dependent CR acceleration. It is this combination of injection
efficiencies obtained from simulation and the theoretical spectral
slope, Eq.~(\ref{eq:qOfM}), which allows us to extend the simulation
spectra far beyond in rigidity that any simulation may possibly reach.
Obviously, this extension is justified because our simulation spectra
reach the asymptotic DSA power-law regimes, Fig.~\ref{fig:energy_dist}.

We focus on the Sedov-Taylor phase of the SNR evolution, during which
the shock radius increases with time as $R_{s}\simeq C_{\mathrm{ST}}\,t^{2/5}$,
while the shock velocity decreases as $V_{s}\simeq(2/5)\,C_{\mathrm{ST}}\,t^{-3/5}=(2/5)\,C_{\mathrm{ST}}^{5/2}\,R_{s}^{-3/2}$,
with $C_{\mathrm{ST}}\simeq(2\,E_{e}/\rho_{0})^{1/5}$. Here $E_{e}$
is the ejecta energy of the supernova and $\rho_{0}$ is the ambient
density. During the increase of the shock radius from $R_{\mathrm{min}}$
to $R_{\mathrm{max}}$, the number of CR particles of species $\alpha$
which are deposited in the shock interior can be calculated as 
\vspace{-5pt}
\begin{equation}
N_{\alpha}(p)\propto\int_{R_{\mathrm{min}}}^{R_{\mathrm{max}}}f_{\alpha}(p,M(R))R^{2}\,\mathrm{d}R\propto\int_{M_{\mathrm{max}}^{-2}}^{M_{\mathrm{min}}^{-2}}f_{\alpha}(p,M)\,\mathrm{d}M^{-2}.\label{eq:number}
\vspace{-3pt}
\end{equation}
Here the spectra are represented in the following way: 
\vspace{-5pt}
\begin{equation}
  f_{\alpha}\propto\eta_{\alpha}(M)(\mathcal{R}/\mathcal{R}_{\mathrm{inj}})^{-q(M)} \qquad \text{with} \quad q(M)=4/(1-M^{-2}).
  \label{eq:spectra}
\vspace{-5pt}
\end{equation}

\begin{wrapfigure}{r}{0.5\columnwidth}%
\vspace{-5pt}
\includegraphics[width=0.48\textwidth]{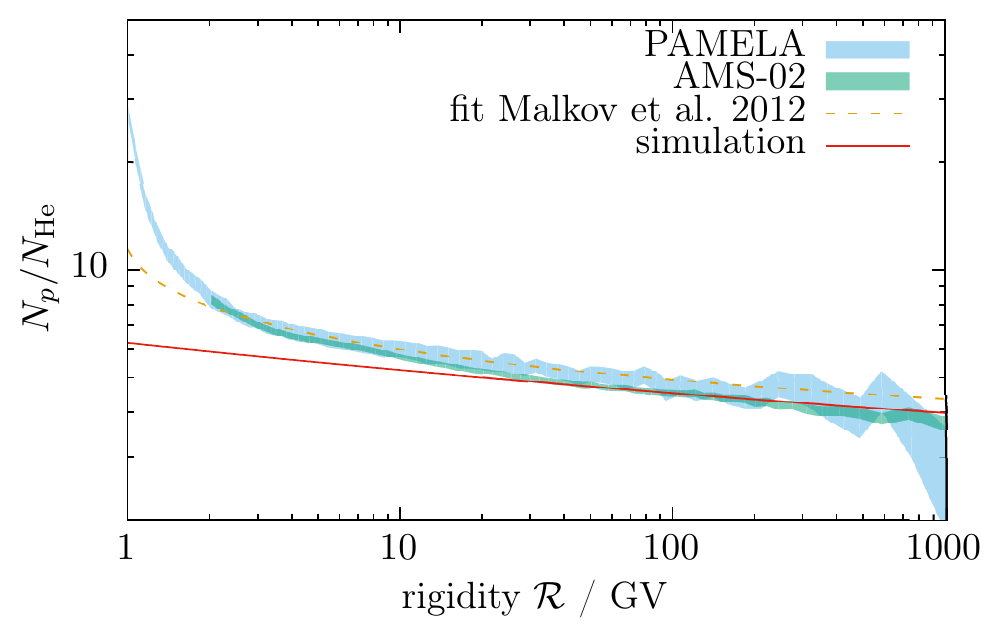}
\vspace{-10pt}
\caption{Proton-to-helium ratio as a function of particle rigidity. The results
from the simulation (red line) are compared to the PAMELA and AMS-02
data. The observed $p/$He ratio is accurately reproduced in the range
$\mathcal{R}\gtrsim10$, as expected theoretically (see text). \label{fig:pHe-ratio}}
\vspace{-5pt}
\end{wrapfigure}

Instead of feeding the simulation data for $\eta_{\alpha}\left(M\right)$
directly to the convolution given by Eqs.~(\ref{eq:number}) and (\ref{eq:spectra})
we take a more practical approach. By fitting the following simple
function $\eta_{\mathrm{inj}}(M)=a\,(M-b)\,M^{-c}$ to the data extracted
from the simulation, we then calculate the proton-to-helium ratio,
$N_{p}/N_{\mathrm{He}}$, according to Eq.~(\ref{eq:number}), as
a function of rigidity. The results are shown in Fig.~\ref{fig:pHe-ratio}
by the red line together with the data from the PAMELA and AMS-02
experiments (blue and green shadow areas). 

Our simulations correctly predict the decrease in proton-to-helium
ratio with increasing rigidity, shown in Fig.~\ref{fig:pHe-ratio},
at exactly the rate measured in the experiments, $\Delta q\approx0.1$,
for $\mathcal{R}\gtrsim10$ GV. Only at lower rigidities, $\mathcal{R}\lesssim10$
GV, the difference between the data and our predictions becomes noticeable.
In fact, some difference has to be expected from Eqs.~(\ref{eq:RigMotion})
and (\ref{eq:CoordMotion}), as the rest mass rigidity $\mathcal{R}_{0}$
is different for $p$ and He by a factor of two. Based on our discussion
in Sec.~\ref{sec:Introduction}, the difference \emph{must} occur because
the equations of motion for $p$ and He deviate towards lower rigidities.
Whether the resulting deviation from our prediction based on Eqs.~(\ref{eq:number})
and (\ref{eq:spectra}) comes from the accelerator(s), propagation
through the ISM or solar modulation, remains unclear. Except for this
uncertainty, the suggested mechanism for $A/Z$-dependence of the
injection fully explains the measured $p$/He ratio.

\acknowledgments
\vspace{-5pt}
The research was supported by DFG grant TL 2479/2-1,
RFBR 16-01-00209, and by NASA ATP-program under grant NNX14AH36G.
The authors acknowledge the North-German Supercomputing Alliance (HLRN) for providing
the computational resources for the simulations.

\end{document}